\documentclass[aps,prx,superscriptaddress,twocolumn]{revtex4}

\usepackage{graphicx}
\usepackage{dcolumn}
\usepackage{bm}
\usepackage{amssymb}
\usepackage{amsfonts}
\usepackage{subfig}
\usepackage{latexsym}
\usepackage{enumerate}
\usepackage{color} 
\usepackage{amsmath}
\usepackage[dvipdfmx]{epsfig}
\usepackage{times}
\usepackage{caption}
\usepackage{fancybox}

\newcommand{\red}{\color{black}}

\newtheorem{theorem}{Theorem}

\newtheorem{conjecture}{Conjecture}

\begin{document}
\widetext
\title{Verification of Many-Qubit States}
YITP-18-59
\author{Yuki Takeuchi}
\email{takeuchi@qi.mp.es.osaka-u.ac.jp}
\affiliation{Graduate School of Engineering Science, Osaka University, 
Toyonaka, Osaka 560-8531, Japan}
\author{Tomoyuki Morimae}
\email{tomoyuki.morimae@yukawa.kyoto-u.ac.jp}
\affiliation{
Department of Computer Science, Gunma University, 1-5-1 Tenjin-cho, 
Kiryu-shi, Gunma 376-0052, Japan}
\affiliation{JST, PRESTO, 4-1-8 Honcho, Kawaguchi, Saitama, 332-0012, Japan}
\affiliation{Yukawa Institute for Theoretical Physics, Kyoto University, 
Kitashirakawa Oiwakecho, Sakyo-ku, Kyoto 606-8502, Japan}

\begin{abstract}
Verification is a task to check whether a given quantum state is close to an ideal state or not. In this paper, we show that a variety of many-qubit
quantum states can be verified with only sequential single-qubit measurements
of Pauli operators. First, we introduce a protocol for
verifying ground states of Hamiltonians. We next explain how to verify quantum states generated by a certain class of quantum circuits. We finally propose an adaptive test of stabilizers that enables the
verification of all polynomial-time-generated
hypergraph states, 
which include output states of the Bremner-Montanaro-Shepherd-type
instantaneous quantum polynomial time (IQP) circuits. 
Importantly, we do not make any assumption that  the identically and independently distributed copies of the same states are given: Our protocols work even if some highly complicated
entanglement is created among copies 
in any artificial way. 
As applications, we consider the verification of the quantum computational supremacy
demonstration with IQP models, and verifiable blind quantum computing.
\end{abstract}

\maketitle
\section{Introduction}
Quantum computing is expected to solve 
several problems exponentially faster than 
classical computing, and therefore, realizing 
universal quantum computers is one of the most central goals
in modern quantum information science.
Output states of even simpler quantum circuits
are also useful.
For example, quantum circuits consisting of only Clifford
gates, which are actually classically simulatable~\cite{[G99]}, can generate
important resources for quantum metrology~\cite{[BIWH96]} and measurement-based
quantum computing (MBQC)~\cite{[RB01]}.
Furthermore, it has recently been shown that output states
of several subuniversal circuits, such as boson sampling, instantaneous quantum polynomial time (IQP), and deterministic quantum computation with one quantum bit (DQC1), 
can generate certain probability distributions that cannot be classically
efficiently sampled unless the polynomial-time hierarchy collapses~\cite{[AA13],[B15],[KL98],[MFF14],[FKMNTT16],[M17],[SB09],[BJS11],[BMS16],[TT16],[GWD17],[VHSRE17],[MSM17],[FU15],[BMZ16],[F16],[HBSE17],[BFK17]}.
Other quantum advantages have also been actively studied~\cite{[BISBDJBMN16],[BGK17]}.
Moreover, ground states of Hamiltonians are important. 
Generating ground states of local Hamiltonians is, in general, quantum Merlin-Arthur (QMA)-hard~\cite{[KSV02]} (which suggests that it is much harder than polynomial-time quantum computing), but several local Hamiltonians
offer important quantum abilities with their
ground states, such as
topologically protected quantum memory~\cite{[K03]}, 
adiabatic quantum computing~\cite{[FGGLLP01]}, 
and MBQC~\cite{[RB01],[GE07],[BM08],[GB08],[DB09],[CDNMB09],[JDBBR09],[M10],[CMDB10],[WAR11],[M11],[LBKRW11],[DBB12],[FM12],[ESBD12],[FNOM13],[MM15]}. In this way, many-qubit quantum states are essential resources for quantum information processing.

\begin{figure}[t]
\begin{center}
\includegraphics[width=8cm, clip]{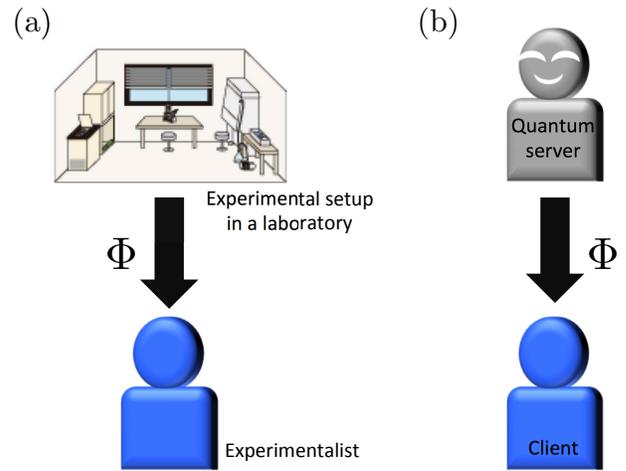}
\end{center}
\captionsetup{justification=raggedright,singlelinecheck=false}
\caption{The prover-verifier game considered in this paper. (a) An experimentalist (verifier) wants to verify the correctness of a state $\Phi$ from the experimental setup (prover). (b) In cloud quantum computing of the type of Ref.~\cite{[MF13]}, a client (verifier) asks a remote server (prover) to generate and send a certain quantum many-qubit state $\Phi$. The client wants to verify the correctness of the state sent from the server.}
\label{setting}
\end{figure}

When an experimentalist generates these many-qubit resource states in his or her own laboratory [Fig.~\ref{setting}(a)], or when a client of cloud quantum computing receives these resource states 
from a remote server [Fig.~\ref{setting}(b)], it is important to 
check the correctness of given states~\cite{FN1}. More precisely, let us consider the following game between
two people, the verifier and the prover~\cite{[FN3]}.
The prover sends a certain quantum state $\Phi$ to the verifier
claiming that it is the tensor product of
many copies $\rho^{\otimes k}$
of a many-qubit state $\rho$.
The state $\rho$ is an important resource state for the verifier.
For example, $\rho$ is a ground state of a Hamiltonian
or a resource state of MBQC.
However, the prover is not necessarily trusted, and therefore,
the verifier has to check the correctness of the given state. 

If $\Phi$ is guaranteed to be at least the tensor product of many copies $\sigma^{\otimes k}$
of the same state $\sigma$, i.e., the states are 
independent and identically distributed (i.i.d.), 
and if the size of $\sigma$ is small, 
the quantum tomography~\cite{[JKMW01]} is enough. 
However, useful resource states
are often large-size quantum states,
and therefore, the quantum tomography suffers from
the exponential blowup.
The exponential increase of parameters is somehow mitigated by using the compressed sensing idea, especially for low-rank quantum states~\cite{GLFBE10}, but the scaling is, in general, exponential. If we are interested only in the fidelity, by direct fidelity estimation~\cite{FL11} and by using fidelity witnesses~\cite{GKEA17}, we achieve the goal without reconstructing the full state, which is more efficient than quantum tomography.

However, these protocols also assume the i.i.d. property of quantum states.
In reality, such an i.i.d. property does not hold.
Because of environmental noises, the generated state in a laboratory
is not a tensor product of the same states.
In cloud quantum computing, moreover, the situation
is worse because a malicious prover might generate highly complicated entanglement
among samples to fool the verifier.
In other words, what the verifier actually receives
is not $\rho^{\otimes k}$ but ${\mathcal E}(\rho^{\otimes k})$
with a completely positive and trace-preserving (CPTP) map ${\mathcal E}$.
If ${\mathcal E}(\rho^{\otimes k})$ is a state generated
by a well-controlled experimental setup,
${\mathcal E}$ is a time
evolution generated by a physically natural Hamiltonian
describing the interaction between the system and
the environment. If
${\mathcal E}(\rho^{\otimes k})$ is a state given
by the server of cloud quantum computing,
${\mathcal E}$ can be any CPTP map~\cite{FN2}.

In addition to the non-i.i.d. property of samples,
another realistic assumption
in verifications is
that the verifier's ability is severely
limited. (In fact, otherwise the verification task would be trivial.
For example, if the verifier can generate the correct state
by his or herself, the verification is straightforward by doing the SWAP test between the given state
and the correct state generated by him or her~\cite{mixed}.)
If the verifier is severely restrictive,
the verification is a highly nontrivial problem.
For example, can the verifier verify a highly entangled 
many-qubit state
by measuring each qubit individually?

In summary, a verification protocol should satisfy
the following three conditions: 
\begin{itemize}
\item[(i)] It runs in polynomial time. 
\item[(ii)] The i.i.d. property of samples is not assumed. 
\item[(iii)] No entangling operation is required for the verifier.
\end{itemize}
Verification protocols that satisfy these three conditions have been 
proposed for some specific classes of states, such as
graph states~\cite{[HM15],[MNS16]} and hypergraph states 
with low connectivity~\cite{[MTH17]}, including the Union Jack state~\cite{[MM16]}.
Here, hypergraph states are generalizations
of graph states by replacing the controlled-$Z$ ($CZ$) gates
of graph states with generalized $CZ$ gates. 
A generalized $CZ$ gate is a unitary gate that 
flips the phase $\pm1$ if and only if all qubits are 
$|1\rangle$.
(See Sec. IV A for the definition of hypergraph states.) 
We say that a hypergraph state has a low connectivity if
the connectivity
\begin{eqnarray}
\label{connectivity}
\xi\equiv\max_{v\in V}\xi_v
\end{eqnarray}
is constant with respect to $|V|$,
where
$\xi_v$ is the number of generalized $CZ$ gates acting
on the vertex $v$, 
$V$ is the set of vertices, and $|V|$ is the size of $V$.

These protocols, Refs.~\cite{[HM15],[MNS16],[MTH17]}, satisfy all the above conditions (i)--(iii). In particular, in these protocols, the verifier only needs sequential
single-qubit measurements of Pauli operators.
However, these protocols leave the following two problems open:
\begin{itemize}
\item[1.]
Are there other more general classes of states that are verifiable
with only sequential single-qubit measurements of Pauli operators?
For example, can we verify ground states of Hamiltonians
and states generated by general quantum circuits
with sequential single-qubit measurements of Pauli operators?
\item[2.]
Can we verify hypergraph states with high connectivity
by using only sequential single-qubit measurements
of Pauli operators?
\end{itemize}
Here, high connectivity means that Eq.~(\ref{connectivity}) is polynomial 
with respect to $|V|$.
The second open problem is important for the verification
of the quantum computational supremacy demonstration because 
output states of the Bremner-Montanaro-Shepherd-type IQP circuits~\cite{[BMS16]}
are hypergraph states with high connectivity.
(See Sec. V for details.)

In this paper, we solve the two open problems by
proposing three verification protocols.
We first introduce a protocol for verifying ground states of Hamiltonians (Sec. II).
We next show a protocol for verifying quantum states generated by 
a certain class of quantum circuits (Sec. III).
As a common technique used in these two verification protocols, we decompose an operator such as a Hamiltonian or a generalized stabilizer into Pauli operators and estimate overlaps between the verified state and Pauli operators. A similar technique was used in the direct fidelity estimation~\cite{FL11}.
We finally explain a verification protocol for hypergraph states with 
high connectivity (Sec. IV). 
For the construction of the third protocol, we propose 
a novel test, which we call the adaptive stabilizer test, by combining 
the stabilizer test of Ref.~\cite{[HM15]} with adaptive classical processing. 
This adaptivity is the key that 
enables the verification of hypergraph states with high 
connectivity.
The previous protocol~\cite{[MTH17]} 
is not enough 
to verify hypergraph states with high
connectivity. The adaptive classical processing we introduce
in Secs. IV B and IV C is the key idea to
realize 
a verification protocol for hypergraph states with high 
connectivity.

The validness of our protocols is demonstrated by showing
their completeness and soundness. Roughly speaking, if the verifier accepts the ideal quantum state with high probability, we say that the verification protocol has the completeness. On the other hand, if the protocol guarantees that a quantum state passing the verification protocol is close to the ideal state with high probability, we say that the protocol has the soundness. The precise statements are given later as theorems.

In Sec.V, we discuss applications of our protocols
to the verification of
quantum computational supremacy demonstrations with the IQP model and its variants. We also consider an application to verifiable blind quantum computing.
Sections VI and VII are devoted to the discussion and the conclusion, respectively.

Note that in addition to Refs.~\cite{[HM15],[MNS16],[MTH17]}
and the present paper, 
other papers have proposed verification protocols for quantum computational supremacy demonstrations. Hangleiter {\it et al.} have proposed a polynomial-time verification protocol for ground states of frustration-free Hamiltonians~\cite{[HKSE17]}. 
A disadvantage of this protocol when it is used for the verification of
quantum computing is that the Feynman-Kitaev history 
state~\cite{[F86],[KSV02]} corresponding to the quantum circuit, 
which is more complicated than the
mere output state of the circuit, has to be generated.
Furthermore, their verification protocol requires multiqubit measurements. 
Based on the verification protocol of Ref.~\cite{[HKSE17]},
Gao {\it et al.}~\cite{[GWD17]} and 
Bermejo-Vega {\it et al.}~\cite{[VHSRE17]} have 
proposed verification protocols for quantum computational supremacy 
demonstrations of their architectures. 
Miller {\it et al.}~\cite{[MSM17]} 
have proposed a polynomial-time verification protocol 
for the output states of the Bremner-Montanaro-Shepherd-type
IQP circuits~\cite{[BMS16]}. 
Their protocol is a special case of our
third protocol when the target state is restricted
to hypergraph states.
With respect to the boson sampling model~\cite{[AA13]}, a verification protocol has already been proposed~\cite{AGKE15}, but this protocol requires at most exponentially many copies of a verified quantum state.
As a common drawback of all these protocols~\cite{[HKSE17],
[MSM17],[GWD17],[VHSRE17],AGKE15}, they
assume the i.i.d. property of samples. 

All verification protocols introduced above and our present protocols require the ability of measurements for the verifier. On the other hand, there are complement protocols where a verifier is required to prepare quantum states~\cite{[KD17],[MPKK17]}. The protocol in Ref.~\cite{[KD17]} uses trap qubits~\cite{[ABE10],[FK17]} to perform verified quantum computational supremacy demonstrations for an Ising sampler~\cite{[GWD17]} or an IQP circuit~\cite{[SB09],[BJS11],[BMS16]}, and does not assume the i.i.d. property. The protocol in Ref.~\cite{[MPKK17]} can verify that the server has the ability to sample from an IQP circuit. For some experimental setups, measurements are easier than  
preparations, and vice versa for other experimental setups.
Therefore, at this moment, we do not know which approach is better.

\section{VERIFICATION OF GROUND STATES of Hamiltonians}
In this section, we explain our verification protocol 
for ground states of Hamiltonians.
In Sec. II A, we define a test. 
In Sec. II B, we explain how to verify ground states by using the test.

\subsection{Test}
Let $H$ be an $N$-qubit Hamiltonian. We want to verify its ground state corresponding to the ground energy $E_0$.
Let $\Delta(>0)$ be a lower bound of the energy gap, i.e., $E_1-E_0\ge\Delta$, where $E_1$ is the first excitation energy. From $H$, we define a rescaled Hamiltonian
\begin{eqnarray}
\label{H'}
H'\equiv\cfrac{H-E_0I^{\otimes N}}{\Delta}.
\end{eqnarray}
Since $H'$ is Hermitian, if we decompose $H'$ in the Pauli basis as
\begin{eqnarray}
\label{H}
H'=\sum_{i=0}^h c_i\tau_i,
\end{eqnarray}
$c_i$ is a real number, where $h=4^N-1$, 
\begin{eqnarray*}
\tau_i\equiv\bigotimes_{j=1}^N\sigma_{i|j},
\end{eqnarray*}
$\sigma_{i|j}\in\{I,X,Y,Z\}$, and $\tau_0\equiv I^{\otimes N}$. From Eq.~(\ref{H'}), the ground energy of $H'$ is $0$. Accordingly,
\begin{eqnarray}
\label{c0}
c_0=c_0\cfrac{{\rm Tr}[I^{\otimes N}]}{2^N}+\sum_{i=1}^h c_i\cfrac{{\rm Tr}[\tau_i]}{2^N}={\rm Tr}\left[H'\cfrac{I^{\otimes N}}{2^N}\right]\ge 0.
\end{eqnarray}
Hereafter, we consider Hamiltonians that satisfy the following three conditions:
\begin{enumerate}
\item[(i)] The probability distribution from $\{|c_i|/R\}_{i=0}^h$ can be sampled exactly in polynomial time. Here, $R\equiv\sum_{i=0}^h|c_i|$.
\item[(ii)] $R=O({\rm poly}(N))$.
\item[(iii)] $R$ is known or can be computed in polynomial time.
\end{enumerate}
Condition (i) is necessary to perform the test defined in the next paragraph. Condition (ii) is required to extract the information of ${\rm Tr}[\rho H']$ from $p_{\rm pass}$ in Eq.~(\ref{p_pass_1_1}) using only the polynomial number of quantum states $\rho$.
Condition (iii) is needed to define the accept or reject criteria in Eq.~(\ref{pass_ground}).
Note that it is obvious that for the usual Hamiltonians in condensed matter physics, such as Ising models and Heisenberg models, these conditions are satisfied if the energy gap is constant or polynomially decays. On the other hand, if the energy gap exponentially decays, then condition (ii) is not satisfied. In fact, for a Hamiltonian $H=\sum_{i=1}^hd_i\tau_i$ with $|d_i|\le{\rm const}$ and $h=O({\rm poly}(N))$,
\begin{eqnarray}
\nonumber
R=\cfrac{\sum_{i=1}^h|d_i|+|E_0|}{\Delta}\ge\cfrac{\sum_{i=1}^h|d_i|}{\Delta}=O(2^{{\rm poly}(N)}).
\end{eqnarray}

The test on an $N$-qubit quantum state $\rho$ is defined as follows: The verifier selects $i$ with probability $|c_i|/R$. If the verifier selects $i$, the verifier measures the $j$th qubit of $\rho$ in the Pauli basis $\sigma_{i|j}$. Let $m_j\in\{1,-1\}$ be the outcome of the measurement on the $j$th qubit. Note that if $\sigma_{i|j}=I$, the verifier sets $m_j=1$. We say that the verifier passes the test on $\rho$ if 
\begin{eqnarray*}
\prod_{j=1}^Nm_j={\rm sgn}(c_i).
\end{eqnarray*}
Here, ${\rm sgn}(\cdot)$ is the sign function.

The expected probability $p_{\rm pass}$ that the verifier passes the test on $\rho$, where the expectation is taken over the sampling of $i$, is
\begin{eqnarray}
\nonumber
p_{\rm pass}&=&\cfrac{|c_0|}{R}+\sum_{i=1}^h\cfrac{|c_i|}{R}{\rm Tr}\left[\rho\cfrac{I^{\otimes N}+{\rm sgn}(c_i)\tau_i}{2}\right]\\
\nonumber
&=&\cfrac{|c_0|}{R}{\rm Tr}\left[\rho\cfrac{I^{\otimes N}+{\rm sgn}(c_0)I^{\otimes N}}{2}\right]\\
\nonumber
&&+\sum_{i=1}^h\cfrac{|c_i|}{R}{\rm Tr}\left[\rho\cfrac{I^{\otimes N}+{\rm sgn}(c_i)\tau_i}{2}\right]\\
\label{p_pass_1_1}
&=&\cfrac{1}{2}+\cfrac{{\rm Tr}\left[\rho H'\right]}{2R},
\end{eqnarray}
where we have used Eqs.~(\ref{c0}) and (\ref{H}) to derive the second and the last equalities, respectively. Note that in order to relate $p_{\rm pass}$ to ${\rm Tr}[\rho H']$, $i=0$ is included in the test.

\begin{figure}[t]
\begin{center}
\includegraphics[width=6cm, clip]{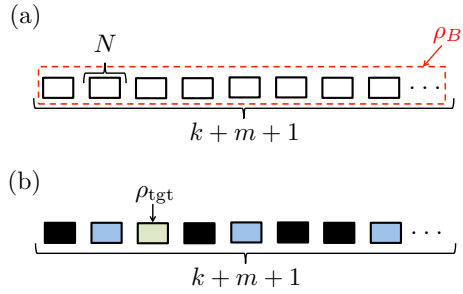}
\end{center}
\captionsetup{justification=raggedright,singlelinecheck=false}
\caption{(a) The quantum state $\rho_B$ in step 1. Each rectangle represents a register that stores $N$ qubits, and $\rho_B$ consists of $k+m+1$ registers. If the prover is honest, the state of each register is the ideal quantum state. On the other hand, if the prover is malicious, registers may be entangled with each other. (b) A quantum state in step 2. Randomly chosen black registers are discarded, and then the remaining $m$ blue registers and one green register become close to i.i.d. samples because of the quantum de Finetti theorem~\cite{[LS15]}. Randomly chosen $m$ blue registers are used for the test. The green register is the target state $\rho_{\rm tgt}$.}
\label{situation}
\end{figure}

\subsection{Verification}
In this subsection, we propose a verification protocol 
for ground states based on the test 
explained in the previous subsection. Our protocol runs as follows:
\begin{enumerate}
\item The prover sends the verifier an $N(k+m+1)$-qubit state $\rho_B$ [see Fig.~\ref{situation} (a)]. The state $\rho_B$ consists of $k+m+1$ registers, and each register stores $N$ qubits. If the prover is honest, the prover sends the tensor product of the ideal state. On the other hand, if the prover is malicious, the prover sends an $N(k+m+1)$-qubit, completely arbitrary, quantum state instead of the tensor product of the ideal state.

\item The verifier chooses $m$ registers uniform randomly and discards them to guarantee that the remaining $N(k+1)$-qubit state $\rho'_B$ is close to an i.i.d. sample by using the quantum de Finetti theorem~\cite{[LS15]}. Next, the verifier chooses one register---which we call the target register, whose state is $\rho_{\rm tgt}$---uniform randomly and uses it for the verifier's purpose. The verifier performs the test on each of the remaining $k$ registers [see Fig.~\ref{situation} (b)]. Let $K_{\rm pass}$ be the number of times that the verifier passes the test. If 
\begin{eqnarray}
\label{pass_ground}
\cfrac{K_{\rm pass}}{k}\le \cfrac{1}{2}+\cfrac{\epsilon}{2R},
\end{eqnarray}
we say that the verifier accepts the prover, where $0<\epsilon<1$ is specified later.
\end{enumerate}
Note that since the random selection is equivalent to random permutation of registers, $\rho_B$ becomes permutation invariant after the random selection in step 2. Accordingly, we can use the quantum de Finetti theorem~\cite{[LS15]}. This idea comes from Ref.~\cite{[MTH17]}.
Hereafter, we consider the case where $\epsilon=1/(4N^2)$, $m\ge2N^5k^2\log{2}$, and $k\ge 32R^2N^5$ are satisfied. In this case, the following theorems hold.
\begin{theorem}[Completeness]
If the prover is honest, i.e., the state of each register is a ground state of $H$, the probability that the verifier accepts the prover is larger than $1-e^{-N}$.
\end{theorem}
{\it Proof.} When the state of each register is a ground state of $H$, $p_{\rm pass}=1/2$. Because of the Hoeffding inequality,
\begin{eqnarray*}
&&{\rm Pr}[{\rm the\ verifier\ accepts\ the\ prover}]\\
&=&1-{\rm Pr}\left[\cfrac{K_{\rm pass}}{k}>\cfrac{1}{2}+\cfrac{\epsilon}{2R}\right]\\
&=&1-{\rm Pr}\left[\cfrac{K_{\rm pass}}{k}>p_{\rm pass}+\cfrac{\epsilon}{2R}\right]\\
&\ge&1-e^{-2\epsilon^2k/(4R^2)}\\
&\ge&1-e^{-N}.
\end{eqnarray*}
\hspace{\fill}$\blacksquare$

\begin{theorem}[Soundness]
If the verifier accepts the prover, the state $\rho_{\rm tgt}$ of the target register satisfies 
\begin{eqnarray*}
{\rm Tr}[\Pi\rho_{\rm tgt}]\ge 1-\cfrac{1}{N}
\end{eqnarray*}
with a probability larger than $1-1/N$. Here, $\Pi$ is the projector onto the ground-energy eigenspace of $H$, and we consider the case where $N\neq 2$.
\end{theorem}
{\it Proof.} Let $\Pi^\perp\equiv I^{\otimes N}-\Pi$, and $T$ be the positive-operator-valued-measure (POVM) element corresponding to the event where the verifier accepts the prover. When $N\neq2$, we can show that for any $N$-qubit state $\rho$,
\begin{eqnarray}
\label{bound1}
{\rm Tr}[(T\otimes\Pi^\perp)\rho^{\otimes k+1}]\le\cfrac{1}{2N^2}.
\end{eqnarray}
Its proof is given later. Because of the quantum de Finetti theorem [for the fully one-way local operations and classical communication (LOCC) norm]~\cite{[LS15]} and Eq.~(\ref{bound1}),
\begin{eqnarray*}
{\rm Tr}[(T\otimes\Pi^\perp)\rho'_B]&\le& {\rm Tr}\left[(T\otimes\Pi^\perp)\int d\mu\rho^{\otimes k+1}\right]\ \ \ \ \ \\
&&+\cfrac{1}{2}\sqrt{\cfrac{2Nk^2\log{2}}{m}}\\
&\le&\cfrac{1}{2N^2}+\cfrac{1}{2N^2}=\cfrac{1}{N^2}.
\end{eqnarray*}
Here, $\mu$ is a probability measure on $\rho$. We have 
\begin{eqnarray*}
{\rm Tr}[(T\otimes\Pi^\perp)\rho'_B]={\rm Tr}[(T\otimes I)\rho'_B]{\rm Tr}[\Pi^\perp\rho_{\rm tgt}].
\end{eqnarray*}
Therefore, if  
\begin{eqnarray*}
{\rm Tr}[\Pi^\perp\rho_{\rm tgt}]>\cfrac{1}{N},
\end{eqnarray*}
then 
\begin{eqnarray*}
{\rm Tr}[(T\otimes I)\rho'_B]<\cfrac{1}{N}.
\end{eqnarray*}
This means that if the verifier accepts the prover, 
\begin{eqnarray*}
{\rm Tr}[\Pi\rho_{\rm tgt}]\ge 1-\cfrac{1}{N}
\end{eqnarray*}
with a probability larger than $1-1/N$.

To complete the proof, we show Eq.~(\ref{bound1}). First, we consider the case where ${\rm Tr}[H'\rho]\le2\epsilon$. Let $\{|E'_i\rangle,E'_i\}_i$ be the set of excited eigenstates of $H'$ and their eigenvalues. Since $E'_i\ge 1$,
\begin{eqnarray*}
{\rm Tr}\left[\Pi^\perp\rho\right]&=&\sum_i\langle E'_i|\rho|E'_i\rangle\\
&\le&\sum_iE'_i\langle E'_i|\rho|E'_i\rangle\\
&=&{\rm Tr}[H'\rho]\le2\epsilon.
\end{eqnarray*}
Therefore,
\begin{eqnarray}
\label{1}
{\rm Tr}[(T\otimes\Pi^\perp)\rho^{\otimes k+1}]={\rm Tr}[T\rho^{\otimes k}]{\rm Tr}[\Pi^\perp\rho]\le \cfrac{1}{2N^2}.
\end{eqnarray}
Next, we consider the case where ${\rm Tr}[\rho H']>2\epsilon$. In this case,
\begin{eqnarray*}
p_{\rm pass}=\cfrac{1}{2}+\cfrac{{\rm Tr}[\rho H']}{2R}>\cfrac{1}{2}+\cfrac{\epsilon}{R}.
\end{eqnarray*}
Therefore, because of the Hoeffding inequality,
\begin{eqnarray*}
{\rm Tr}[T\rho^{\otimes k}]&=&{\rm Pr}\left[\cfrac{K_{\rm pass}}{k}\le\cfrac{1}{2}+\cfrac{\epsilon}{2R}\right]\\
&\le&{\rm Pr}\left[\cfrac{K_{\rm pass}}{k}<p_{\rm pass}-\cfrac{\epsilon}{2R}\right]\\
&\le&e^{-2\epsilon^2k/(4R^2)}\\
&\le&e^{-N}.
\end{eqnarray*}
Hence, 
\begin{eqnarray}
\label{2}
{\rm Tr}[(T\otimes\Pi^\perp)\rho^{\otimes k+1}]={\rm Tr}[T\rho^{\otimes k}]{\rm Tr}[\Pi^\perp\rho]\le e^{-N}.
\end{eqnarray}
From Eqs.~(\ref{1}) and (\ref{2}), when $N\neq 2$,
\begin{eqnarray*}
{\rm Tr}[(T\otimes\Pi^\perp)\rho^{\otimes k+1}]&\le&{\rm max}\left(\cfrac{1}{2N^2},e^{-N}\right)\\
&=&\cfrac{1}{2N^2}.
\end{eqnarray*}
\hspace{\fill}$\blacksquare$

\section{Verification of quantum states generated by a certain class of quantum circuits}
In this section, we explain our second verification protocol,
namely, the protocol for quantum states generated by a certain class of quantum circuits. 
In Sec. III A, we explain a stabilizer test. In Sec. III B, we show the verification
protocol based on the stabilizer test.
\subsection{Stabilizer test}
Let us assume that we want to verify the quantum state $|\psi\rangle\equiv U|+\rangle^{\otimes N}$, where $|+\rangle\equiv(|0\rangle+|1\rangle)/\sqrt{2}$, and $U$ is a certain $N$-qubit unitary operator whose properties are specified later. 
The $i$th stabilizer $g_i$ of $|\psi\rangle$ is defined by 
\begin{eqnarray}
\label{U}
g_i\equiv UX_iU^\dag, 
\end{eqnarray}
where $X_i$ is performed on the $i$th qubit of $|\psi\rangle$. Note that $g_i$ is not necessarily a tensor product of Pauli operators, and therefore, it should be considered as a ``generalized stabilizer." From Eq.~(\ref{U}), 
\begin{eqnarray*}
\prod_{i=1}^N\cfrac{I^{\otimes N}+g_i}{2}=|\psi\rangle\langle\psi|.
\end{eqnarray*}
Since $g_i$ is Hermitian, if we decompose $g_i$ in the Pauli basis as
\begin{eqnarray*}
g_i=\sum_{j}c_j^{(i)}\tau_j,
\end{eqnarray*}
$c_j^{(i)}$ is a real number, where 
\begin{eqnarray*}
\tau_j\equiv\bigotimes_{k=1}^N\sigma_{j|k},
\end{eqnarray*}
and $\sigma_{j|k}\in\{I,X,Y,Z\}$. Hereafter, we consider the $U$ that satisfies the following three conditions:
\begin{enumerate}
\item[(i)] The probability distribution from $\{|c_j^{(i)}|/R_i\}_j$ can be sampled exactly in polynomial time. Here, $R_i\equiv\sum_j|c_j^{(i)}|$.
\item[(ii)] $R\equiv{\rm max}(R_1,\cdot\cdot\cdot,R_N)=O({\rm poly}(N))$.
\item[(iii)] $R_i$ is known or can be computed in polynomial time for all $i$.
\end{enumerate}
Condition (i) is necessary to perform the stabilizer test defined in the next paragraph. Condition (ii) is required to extract the information of ${\rm Tr}[\rho g_i]$ from $p_{\rm pass}(i)$ in Eq.~(\ref{p_pass_2}) using only the polynomial number of quantum states $\rho$. Condition (iii) is needed to define the accept or reject criteria in Eq.~(\ref{pass_circuit}).

The stabilizer test for $g_i$ on $\rho$ is defined as follows: The verifier selects $j$ with probability $|c_j^{(i)}|/R_i$. The verifier measures the $k$th qubit of $\rho$ in the Pauli basis $\sigma_{j|k}$. Let $m_k\in\{1,-1\}$ be the outcome of the measurement on the $k$th qubit. Note that if $\sigma_{j|k}=I$, the verifier sets $m_k=1$. We say that the verifier passes the stabilizer test for $g_i$ on $\rho$ if 
\begin{eqnarray*}
\prod_{k=1}^Nm_k={\rm sgn}(c_j^{(i)}).
\end{eqnarray*}
Since quantum states satisfying the above three properties include graph states and hypergraph states with low connectivity as special cases, our stabilizer test can be considered as a generalization of previous stabilizer tests~\cite{[HM15],[MTH17]}.

The expected probability $p_{\rm pass}(i)$ that the verifier passes the stabilizer test for $g_i$ on $\rho$, where the expectation is taken over the sampling of $j$, is
\begin{eqnarray}
\nonumber
p_{\rm pass}(i)&=&\sum_j\cfrac{|c_j^{(i)}|}{R_i}{\rm Tr}\left[\rho\cfrac{I^{\otimes N}+{\rm sgn}(c_j^{(i)})\tau_j}{2}\right]\\
\label{p_pass_2}
&=&\cfrac{1}{2}+\cfrac{{\rm Tr}[\rho g_i]}{2R_i}.
\end{eqnarray}

\subsection{Verification}

In this subsection, we propose a verification protocol for $|\psi\rangle=U|+\rangle^{\otimes N}$. Our protocol runs as follows:
\begin{enumerate}
\item The prover sends the verifier an $N(Nk+m+1)$-qubit state $\rho_B$. The state $\rho_B$ consists of $Nk+m+1$ registers, and each register stores $N$ qubits. If the prover is honest, the prover sends $|\psi\rangle^{\otimes Nk+m+1}$. On the other hand, if the prover is malicious, the prover sends an $N(Nk+m+1)$-qubit, completely arbitrary, quantum state instead of $|\psi\rangle^{\otimes Nk+m+1}$.

\item The verifier chooses $m$ registers uniform randomly and discards them to guarantee that the remaining $N(Nk+1)$-qubit state $\rho'_B$ is close to an i.i.d. sample by using the quantum de Finetti theorem~\cite{[LS15]}. Next, the verifier chooses one register---which we call the target register, whose state is $\rho_{\rm tgt}$---uniform randomly and uses it for the verifier's purpose. The remaining $Nk$ registers are divided into $N$ groups such that which register is assigned to the $i$th group is uniformly random. The verifier performs the stabilizer test for $g_i$ on every register in the $i$th group. Let $K_i$ be the number of times that the verifier passes the stabilizer test for $g_i$. If 
\begin{eqnarray}
\label{pass_circuit}
\cfrac{K_i}{k}\ge \cfrac{1}{2}+\cfrac{1-\epsilon}{2R_i},
\end{eqnarray}
we say that the verifier passes the stabilizer test for the $i$th group, where $0<\epsilon<1$ is specified later. If the verifier passes the stabilizer test for all $i$, we say that the verifier accepts the prover. 
\end{enumerate}
Hereafter, we consider the case where $\epsilon=1/(2N^3)$, $m\ge2N^7k^2\log{2}$, and $k\ge 8R^2N^7$ are satisfied. In this case, the following theorems hold.
\begin{theorem}[Completeness]
If the prover is honest, i.e., the state of each register is $|\psi\rangle$, the probability that the verifier accepts the prover is larger than $1-Ne^{-N}$.
\end{theorem}
{\it Proof.} When the state of each register is $|\psi\rangle$, 
\begin{eqnarray*}
p_{\rm pass}(i)=\cfrac{1}{2}+\cfrac{1}{2R_i}.
\end{eqnarray*}
Because of the union bound and the Hoeffding inequality,
\begin{eqnarray*}
&&{\rm Pr}[{\rm the\ verifier\ accepts\ the\ prover}]\\
&=&{\rm Pr}\left[\bigwedge_{i=1}^N\left(\cfrac{K_i}{k}\ge\cfrac{1}{2}+\cfrac{1-\epsilon}{2R_i}\right)\right]\\
&\ge&1-\sum_{i=1}^N{\rm Pr}\left[\cfrac{K_i}{k}<p_{\rm pass}(i)-\cfrac{\epsilon}{2R_i}\right]\ \ \ \ \ \ \ \\
&\ge&1-\sum_{i=1}^Ne^{-2\epsilon^2k/(4R_i^2)}\\
&\ge&1-Ne^{-2\epsilon^2k/(4R^2)}\\
&\ge&1-Ne^{-N}.
\end{eqnarray*}
\hspace{\fill}$\blacksquare$

\begin{theorem}[Soundness]
If the verifier accepts the prover, the state $\rho_{\rm tgt}$ of the target register satisfies 
\begin{eqnarray*}
\langle\psi|\rho_{\rm tgt}|\psi\rangle\ge 1-\cfrac{1}{N}
\end{eqnarray*}
with a probability larger than $1-1/N$. Here, we consider the case where $N\neq 2$.
\end{theorem}
{\it Proof.} Let $\Pi^\perp$ be the $N$-qubit projector $I^{\otimes N}-|\psi\rangle\langle\psi|$, and $T$ be the POVM element corresponding to the event where the verifier accepts the prover. When $N\neq2$, we can show that for any $N$-qubit state $\rho$,
\begin{eqnarray}
\label{bound2}
{\rm Tr}[(T\otimes\Pi^\perp)\rho^{\otimes Nk+1}]\le\cfrac{1}{2N^2}.
\end{eqnarray}
Its proof is given later. Because of the quantum de Finetti theorem (for the fully one-way LOCC norm)~\cite{[LS15]} and Eq.~(\ref{bound2}),
\begin{eqnarray*}
{\rm Tr}[(T\otimes\Pi^\perp)\rho'_B]&\le& {\rm Tr}\left[(T\otimes\Pi^\perp)\int d\mu\rho^{\otimes Nk+1}\right]\ \ \ \ \ \\
&&+\cfrac{1}{2}\sqrt{\cfrac{2N^3k^2\log{2}}{m}}\\
&\le&\cfrac{1}{2N^2}+\cfrac{1}{2N^2}=\cfrac{1}{N^2}.
\end{eqnarray*}
Here, $\mu$ is a probability measure on $\rho$. We have 
\begin{eqnarray*}
{\rm Tr}[(T\otimes\Pi^\perp)\rho'_B]={\rm Tr}[(T\otimes I)\rho'_B]{\rm Tr}[\Pi^\perp\rho_{\rm tgt}].
\end{eqnarray*}
Therefore, if 
\begin{eqnarray*}
{\rm Tr}[\Pi^\perp\rho_{\rm tgt}]>\cfrac{1}{N},
\end{eqnarray*}
then 
\begin{eqnarray*}
{\rm Tr}[(T\otimes I)\rho'_B]<\cfrac{1}{N}.
\end{eqnarray*}
This means that if the verifier accepts the prover, 
\begin{eqnarray*}
\langle\psi|\rho_{\rm tgt}|\psi\rangle\ge 1-\cfrac{1}{N}
\end{eqnarray*}
with a probability larger than $1-1/N$. 

To complete the proof, we show Eq.~(\ref{bound2}). First, we consider the case where ${\rm Tr}[g_i\rho]\ge1-2\epsilon$ is satisfied for all $i$. From the union bound,
\begin{eqnarray*}
1-\langle\psi|\rho|\psi\rangle&=&1-{\rm Tr}\left[\rho\prod_{i=1}^N\cfrac{I^{\otimes N}+g_i}{2}\right]\\
&\le&\sum_{i=1}^N\left(1-{\rm Tr}\left[\rho\cfrac{I^{\otimes N}+g_i}{2}\right]\right)\\
&\le&N\epsilon.
\end{eqnarray*}
Therefore,
\begin{eqnarray}
\nonumber
{\rm Tr}[(T\otimes\Pi^\perp)\rho^{\otimes Nk+1}]&=&{\rm Tr}[T\rho^{\otimes Nk}]{\rm Tr}[\Pi^\perp\rho]\\
\label{I}
&\le& \cfrac{1}{2N^2}.
\end{eqnarray}
Next, we consider the case where ${\rm Tr}[g_i\rho]<1-2\epsilon$ is satisfied for at least one $i$. In this case, for the $i'$ that satisfies ${\rm Tr}[g_{i'}\rho]<1-2\epsilon$,
\begin{eqnarray*}
p_{\rm pass}(i')=\cfrac{1}{2}+\cfrac{{\rm Tr}[g_{i'}\rho]}{2R_{i'}}<\cfrac{1}{2}+\cfrac{1-2\epsilon}{2R_{i'}}.
\end{eqnarray*}
Therefore, because of the Hoeffding inequality,
\begin{eqnarray*}
{\rm Tr}[T\rho^{\otimes Nk}]&\le&{\rm Pr}\left[\cfrac{K_{i'}}{k}\ge\cfrac{1}{2}+\cfrac{1-\epsilon}{2R_{i'}}\right]\\
&\le&{\rm Pr}\left[\cfrac{K_{i'}}{k}>p_{\rm pass}(i')+\cfrac{\epsilon}{2R_{i'}}\right]\\
&\le&e^{-2\epsilon^2k/(4R_{i'}^2)}\\
&\le&e^{-2\epsilon^2k/(4R^2)}\\
&\le&e^{-N}.
\end{eqnarray*}
Hence, 
\begin{eqnarray}
\nonumber
{\rm Tr}[(T\otimes\Pi^\perp)\rho^{\otimes Nk+1}]&=&{\rm Tr}[T\rho^{\otimes Nk}]{\rm Tr}[\Pi^\perp\rho]\\
\label{II}
&\le&e^{-N}.
\end{eqnarray}
From Eqs.~(\ref{I}) and (\ref{II}), when $N\neq 2$,
\begin{eqnarray*}
{\rm Tr}[(T\otimes\Pi^\perp)\rho^{\otimes Nk+1}]&\le&{\rm max}\left(\cfrac{1}{2N^2},e^{-N}\right)\\
&=&\cfrac{1}{2N^2}.
\end{eqnarray*} \hspace{\fill}$\blacksquare$

\section{VERIFICATION OF HYPERGRAPH STATES}
Although the verification protocol proposed in Sec. III can verify hypergraph states with low connectivity, it cannot verify hypergraph states with high connectivity. To verify hypergraph states with high connectivity, we now explain our third
protocol, which uses a new adaptive stabilizer test. In Sec. IV A, we review the definition of hypergraph states. In Sec. IV B, we explain our basic idea with a simple example. In Sec. IV C, we define the adaptive stabilizer test in a general form. In Sec. IV D, we explain how to verify hypergraph states with
high connectivity by using the adaptive stabilizer test.
\subsection{Hypergraph states}

In this subsection, we review the definition of hypergraph states~\cite{[RHBM13]} and their properties. A hypergraph $G\equiv(V,E)$ is a pair of a set $V$ of vertices and a set $E$ of hyperedges, where a hyperedge is a set of vertices. We define $N\equiv|V|$. The hypergraph state $|G\rangle$ corresponding to the hypergraph $G$ is defined by
\begin{eqnarray*}
|G\rangle\equiv\left(\prod_{e\in E}\widetilde{CZ}_e\right)|+\rangle^{\otimes N},
\end{eqnarray*}
where 
$\widetilde{CZ}_e$ is the generalized $CZ$ gate acting on vertices in $e$; i.e., it is the gate that flips the phase $\pm 1$ if all qubits in $e$ are $|1\rangle$. Since a hypergraph has at most $2^N-1$ hyperedges, the time required to generate a hypergraph state is at most $O(2^N)$. However, in many quantum-information processing protocols, only quantum states that can be generated in polynomial time are used. To focus on such efficiently generatable quantum states, we assume that $2\le|e|\le c$ for all $e\in E$, where $|e|$ is the size of $e$. Here, $c(\ge 3)$ is a constant integer. The size $|E|$ of $E$ is then polynomially upper bounded: 
\begin{eqnarray*}
|E|\le\sum_{k=2}^c\binom{N}{k}=O(N^c).
\end{eqnarray*}

The $i$th stabilizer $g_i$ of $|G\rangle$ ($i=1,2,\ldots,N$) is defined by
\begin{eqnarray}
\label{g_i1}
g_i&\equiv&\left(\prod_{e\in E}\widetilde{CZ}_e\right)X_i\left(\prod_{e\in E}\widetilde{CZ}_e\right),
\end{eqnarray}
where $X_i$ is the Pauli-$X$ operator acting on the $i$th qubit.
We can show the useful relation
\begin{eqnarray*}
\prod_{i=1}^N\cfrac{I^{\otimes N}+g_i}{2}=|G\rangle\langle G|,
\end{eqnarray*}
which is derived by applying 
$\prod_{e\in E}\widetilde{CZ}_e$ from both the right and the left
on both sides of the trivial equation
\begin{eqnarray*}
\prod_{i=1}^N\frac{I+X_i}{2}
=|+\rangle\langle +|^{\otimes N},
\end{eqnarray*}
and by using Eq.~(\ref{g_i1}).

\subsection{Simple example}
Before introducing the general formalism of our adaptive stabilizer test, here we briefly explain our basic idea with a simple example. Let us consider the three-qubit hypergraph state, 
\begin{eqnarray*}
|G\rangle&=&\widetilde{CZ}_{1,2,3}|+\rangle^{\otimes 3}\\
&=&\cfrac{(|00\rangle+|01\rangle+|10\rangle)_{1,2}|+\rangle_3+|11\rangle_{1,2}|-\rangle_3}{2},
\end{eqnarray*}
where $|-\rangle\equiv(|0\rangle-|1\rangle)/\sqrt{2}$. From the definition Eq.~(\ref{g_i1}),
its stabilizers are
calculated as 
\begin{eqnarray*}
g_1&=&\sum_{a\in\{0,1\}}X\otimes|a\rangle\langle a|\otimes Z^a,\\
g_2&=&\sum_{a\in\{0,1\}}|a\rangle\langle a|\otimes X\otimes Z^a,\\
g_3&=&\sum_{a\in\{0,1\}}|a\rangle\langle a|\otimes Z^a\otimes X.
\end{eqnarray*}

\begin{figure}[t]
\begin{center}
\includegraphics[width=6cm, clip]{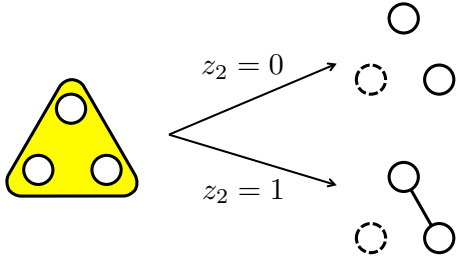}
\end{center}
\captionsetup{justification=raggedright,singlelinecheck=false}
\caption{A quantum state after the $Z_2$-basis measurement on $|G\rangle$. Solid-line and dashed-line circles represent $|+\rangle$ and a measured qubit, respectively. The yellow triangle and the black solid line represent $\widetilde{CZ}_{1,2,3}$ and $\widetilde{CZ}_{1,3}$, respectively.}
\label{ada}
\end{figure}

The adaptive stabilizer test for $g_1$ on 
a three-qubit quantum state $\rho$ proceeds as follows: 
\begin{itemize}
\item[1.]
The verifier measures the first qubit of $\rho$ in the $X$ basis. 
Let $x_1\in\{1,-1\}$ be the measurement result.
\item[2.]
The verifier measures the second and third qubits in the $Z$ bases. 
Let $z_j\in\{1,-1\}$ $(j=2,3)$ be the measurement result
for the $j$th qubit.
\item[3.]
If $z_2=1$ and $x_1=1$, the verifier accepts.
If $z_2=-1$ and $x_1z_3=1$, the verifier accepts.
Otherwise, the verifier rejects.
\end{itemize}
It is easy to check that the acceptance probability $p_{\rm pass}$ of this test
is
\begin{eqnarray*}
p_{\rm pass}={\rm Tr}\left[\rho\frac{I^{\otimes 3}+g_1}{2}\right].
\end{eqnarray*}
(The intuitive idea is illustrated in Fig.~\ref{ada}.
When $z_2=1$,
the three-qubit hypergraph state 
$|G\rangle$ becomes $|+\rangle_1\otimes|+\rangle_3$.
When $z_2=-1$, it becomes the two-qubit graph state 
that is stabilized by $X_1Z_3$.) 
Therefore, the correct state $|G\rangle$
passes the test with probability 1. 
We will see later that 
the estimation of $p_{\rm pass}$ (and therefore the
estimation of ${\rm Tr}[g_1\rho]$) is possible in our verification
protocol. With respect to $g_2$ and $g_3$, a similar argument holds.

\subsection{Adaptive stabilizer test}
For general polynomial-time-generated hypergraph states, we define the adaptive stabilizer test for $g_i$ using the idea explained above. Let $V=\{v_1,v_2,\ldots,v_N\}$ and $E=\{e_1,e_2,\ldots,e_{|E|}\}$. The generalized $CZ$ gate acting on vertices $\{v_1^{(j)},v_2^{(j)},\ldots,v_{|e_j|}^{(j)}\}$ that are connected by the $j$th hyperedge $e_j$ can be written as
\begin{eqnarray}
\nonumber
\widetilde{CZ}_{e_j}&\equiv&\left(\prod_{k=1}^{|e_j|-1}I_{v_k^{(j)}}-\prod_{k=1}^{|e_j|-1}{\ovalbox{1}}_{v_k^{(j)}}\right)I_{v_{|e_j|}^{(j)}}\\
\label{GCZ}
&&+\left(\prod_{k=1}^{|e_j|-1}{\ovalbox{1}}_{v_k^{(j)}}\right)Z_{v_{|e_j|}^{(j)}},
\end{eqnarray}
where $\ovalbox{$a$}\equiv|a\rangle\langle a|$ $(a\in\{0,1\})$.
From Eq.~(\ref{g_i1}), the $i$th stabilizer $g_i$ of $|G\rangle$ can be calculated as
\begin{eqnarray}
\label{g_i}
g_i=X_{v_i}\left(\prod_{v_{i'}\in W_Z^{(i)}}Z_{v_{i'}}\right)\left(\prod_{\tilde{{\bf v}}^{(j)}\in W_{CZ}^{(i)}}\widetilde{CZ}_{\tilde{\bf v}^{(j)}}\right),
\end{eqnarray}
where 
\begin{eqnarray*}
W_Z^{(i)}&\equiv&\{v_{i'}\in V|(v_i,v_{i'})\in E\},\\
W_{CZ}^{(i)}&\equiv&\cup_{e_j\in E'}W_{CZ}^{(i,j)},\\
W_{CZ}^{(i,j)}&\equiv&\{\tilde{{\bf v}}^{(j)}|(v_i,\tilde{{\bf v}}^{(j)})=e_j\}.
\end{eqnarray*}
Here, $E'(\subseteq E)$ is a set of hyperedges that connect more than two vertices, and $\tilde{\bf v}^{(j)}\equiv\tilde{v}_1^{(j)},\ldots,\tilde{v}_{|e_j|-1}^{(j)}$ is a shorthand notation.
By substituting Eq.~(\ref{GCZ}) into Eq.~(\ref{g_i}),
\begin{widetext}
\begin{eqnarray}
\nonumber
g_i&=&X_{v_i}\left(\prod_{v_{i'}\in W_Z^{(i)}}Z_{v_{i'}}\right)\left[\prod_{\tilde{\bf v}^{(j)}\in W_{CZ}^{(i)}}\sum_{{\bf a}^{(j)}\in\{0,1\}^{|e_j|-2}}\left(\prod_{k=1}^{|e_j|-2}{\ovalbox{$a_{\tilde{v}_k^{(j)}}$}}_{\tilde{v}_k^{(j)}}\right)Z_{\tilde{v}_{|e_j|-1}^{(j)}}^{f({\bf a}^{(j)}\cup\{a_{\tilde{v}_{|e_j|-1}^{(j)}}\})}\right]\\
\label{g}
&=&\sum_{{\bf a}\in\{0,1\}^{|{\tilde{W}}_P^{(i)}|}}(-1)^{\alpha^{(i,{\bf a})}}X_{v_i}\left(\prod_{v_{i'}\in {\tilde{W}}_Z^{(i,{\bf a})}}Z_{v_{i'}}\right)\left(\prod_{v_{i'}\in {\tilde{W}}_P^{(i)}}{\ovalbox{$a_{v_{i'}}$}}_{v_{i'}}\right),
\end{eqnarray}
where
\begin{eqnarray}
\label{D'}
\tilde{W}_P^{(i)}&\equiv&\cup_{\tilde{\bf v}^{(j)}\in W_{CZ}^{(i)}}\{\tilde{\bf v}^{(j)}\}\backslash\{\tilde{v}_{|e_j|-1}^{(j)}\},\\
\nonumber
&&(-1)^{\alpha^{(i,{\bf a})}}\left(\prod_{v_{i'}\in {\tilde{W}}_Z^{(i,{\bf a})}}Z_{v_{i'}}\right)\left(\prod_{v_{i'}\in {\tilde{W}}_P^{(i)}}\ovalbox{${a_{v_{i'}}}$}_{v_{i'}}\right)\\
\label{D_a}
&\equiv&\left(\prod_{v_{i'}\in W_Z^{(i)}}Z_{v_{i'}}\right)\left[\prod_{\tilde{\bf v}^{(j)}\in W_{CZ}^{(i)}}\sum_{{\bf a}^{(j)}\in\{0,1\}^{|e_j|-2}}\left(\prod_{k=1}^{|e_j|-2}{\ovalbox{$a_{\tilde{v}_k^{(j)}}$}}_{\tilde{v}_k^{(j)}}\right)Z_{\tilde{v}_{|e_j|-1}^{(j)}}^{f({\bf a}^{(j)}\cup\{a_{\tilde{v}_{|e_j|-1}^{(j)}}\})}\right]\left(\prod_{v_{i'}\in {\tilde{W}}_P^{(i)}}{\ovalbox{$a_{v_{i'}}$}}_{v_{i'}}\right).\ \ \ \ \ \ \
\end{eqnarray}
\end{widetext}
Here, ${\bf a}^{(j)}\equiv\{a_{\tilde{v}_1^{(j)}},\cdot\cdot\cdot,a_{\tilde{v}_{|e_j|-2}^{(j)}}\}$, ${\bf a}\equiv\cup_{\tilde{\bf v}^{(j)}\in W_{CZ}^{(i)}}{\bf a}^{(j)}$, $\alpha^{(i,{\bf a})}\in\{0,1\}$, and $f({\bf a}^{(j)}\cup\{a_{\tilde{v}_{|e_j|-1}^{(j)}}\})$ is a function, where it is equal to $1$ if and only if all elements of $\{a_{\tilde{v}_1^{(j)}},\cdot\cdot\cdot,a_{\tilde{v}_{|e_j|-1}^{(j)}}\}$ are $1$, and it is equal to $0$ in other cases. Note that $\alpha^{(i,{\bf a})}$ and $\tilde{W}_Z^{(i,{\bf a})}$ are defined by Eq.~(\ref{D_a}), and $\tilde{W}_Z^{(i,{\bf a})}\cap\tilde{W}_P^{(i)}=\emptyset$. From Eq.~(\ref{D'}), the time required to derive $\tilde{W}_P^{(i)}$ is at most $O(N^{c-1})$. When the values of all elements of ${\bf a}$ are given and $\tilde{W}_P^{(i)}$ is known, the time required to calculate the rhs of Eq.~(\ref{D_a}) is at most $O(N^{c-1})$. Accordingly, we can derive $\alpha^{(i,{\bf a})}$, $\tilde{W}_Z^{(i,{\bf a})}$, and $\tilde{W}_P^{(i)}$ in classical polynomial time.

The adaptive stabilizer test for $g_i$ on an $N$-qubit quantum state $\rho$ is defined as follows: The verifier measures the $i$th qubit of $\rho$ that corresponds to the vertex $v_i$ in the $X$ basis, and each of the other qubits of $\rho$ in the $Z$ basis, respectively. Let $x_i\in\{1,-1\}$ be the outcome of the $X$-basis measurement, and $z_{i'}\in\{1,-1\}$ be the outcome of the $Z$-basis measurement on the $i'$th qubit. Then, the verifier calculates $\tilde{W}_P^{(i)}$. From $\tilde{W}_P^{(i)}$ and measurement outcomes, the verifier knows the values of ${\bf a}$. We say that the verifier passes the adaptive stabilizer test for $g_i$ on $\rho$ if 
\begin{eqnarray*}
(-1)^{\alpha^{(i,{\bf a})}}x_i\prod_{v_{i'}\in\tilde{W}_Z^{(i,{\bf a})}}z_{i'}=1.
\end{eqnarray*}
Note that the adaptiveness is not needed in the special case of graph states because $W_{CZ}^{(i)}=\emptyset$.

The expected probability $p_{\rm pass}(i)$ that the verifier passes the adaptive stabilizer test for $g_i$ on $\rho$, where the expectation is taken over the sampling of ${\bf a}$, is
\begin{eqnarray}
\nonumber
&&p_{\rm pass}(i)\\
\nonumber
&=&\sum_{{\bf a}:p({\bf a})\neq 0}p({\bf a}){\rm Tr}\left[\cfrac{P^{(i,{\bf a})}\rho P^{(i,{\bf a})}}{p({\bf a})}\cfrac{I^{\otimes |\tilde{W}_Z^{(i,{\bf a})}|+1}+S^{(i,{\bf a})}}{2}\right]\ \ \ \ \ \ \ \\
\nonumber
&=&\cfrac{1}{2}\sum_{{\bf a}:p({\bf a})\neq 0}\left({\rm Tr}\left[\rho P^{(i,{\bf a})}\right]+{\rm Tr}\left[\rho P^{(i,{\bf a})}S^{(i,{\bf a})}\right]\right)\\
\nonumber
&=&\cfrac{1}{2}\sum_{{\bf a}\in\{0,1\}^{|\tilde{W}_P^{(i)}|}}\left({\rm Tr}\left[\rho P^{(i,{\bf a})}\right]+{\rm Tr}\left[\rho P^{(i,{\bf a})}S^{(i,{\bf a})}\right]\right)\\
\nonumber
&=&\cfrac{1}{2}\left(1+{\rm Tr}\left[\rho\sum_{{\bf a}\in\{0,1\}^{|\tilde{W}_P^{(i)}|}}P^{(i,{\bf a})}S^{(i,{\bf a})}\right]\right)\\
\label{16}
&=&\cfrac{1}{2}\left(1+{\rm Tr}[\rho g_i]\right),
\end{eqnarray}
where
\begin{eqnarray*}
p({\bf a})&\equiv&{\rm Tr}\left[\rho P^{(i,{\bf a})}\right], \\
P^{(i,{\bf a})}&\equiv&\prod_{v_{i'}\in\tilde{W}_P^{(i)}}\ovalbox{$a_{v_{i'}}$}_{v_{i'}},\\
S^{(i,{\bf a})}&\equiv&(-1)^{\alpha^{(i,{\bf a})}}X_{v_i}\left(\prod_{v_{i'}\in\tilde{W}_Z^{(i,{\bf a})}}Z_{v_{i'}}\right).
\end{eqnarray*}
We have used Eq.~(\ref{g}) to derive the last equality. 

Let us explain why our adaptive stabilizer test can verify
hypergraph states with high connectivity, while
our second protocol (Sec. III) and the previous protocol~\cite{[MTH17]} cannot.
For the nonadaptive stabilizer test (Sec. III),
the probability $p_{\rm pass}(i)$ of passing the stabilizer test for $g_i$
is given in Eq.~(\ref{p_pass_2}). 
If $R_i=O(\exp({N}))$, exponentially many tests are required
to distinguish $p_{\rm pass}(i)$ from 1/2,
which means that no polynomial-time verification is possible.
On the other hand, since $R_i$ does not appear in Eq.~(\ref{16}), 
such a problem does not occur for the adaptive stabilizer test.

\subsection{Verification}

In this subsection, we propose a verification protocol for hypergraph states based on the adaptive stabilizer test explained in the previous subsection. Our protocol runs as follows:
\begin{enumerate}
\item The prover sends the verifier an $N(Nk+m+1)$-qubit state $\rho_B$. The state $\rho_B$ consists of $Nk+m+1$ registers, and each register stores $N$ qubits. If the prover is honest, the prover sends $|G\rangle^{\otimes Nk+m+1}$. On the other hand, if the prover is malicious, the prover sends an $N(Nk+m+1)$-qubit, completely arbitrary, quantum state instead of $|G\rangle^{\otimes Nk+m+1}$.

\item The verifier chooses $m$ registers uniform randomly and discards them to guarantee that the remaining $N(Nk+1)$-qubit state $\rho'_B$ is close to an i.i.d. sample by using the quantum de Finetti theorem~\cite{[LS15]}. Next, the verifier chooses one register---which we call the target register, whose state is $\rho_{\rm tgt}$---uniform randomly and uses it for the verifier's purpose. The remaining $Nk$ registers are divided into $N$ groups such that which register is assigned to the $i$th group is uniformly random. The verifier performs the adaptive stabilizer test for $g_i$ on every register in the $i$th group. Let $K_i$ be the number of times that the verifier passes the adaptive stabilizer test for $g_i$. If 
\begin{eqnarray}
\label{pass_hyper}
\cfrac{K_i}{k}\ge 1-\epsilon,
\end{eqnarray}
we say that the verifier passes the adaptive stabilizer test for the $i$th group, where $0<\epsilon<1$ is specified later. If the verifier passes the adaptive stabilizer test for all $i$, we say that the verifier accepts the prover. 
\end{enumerate}
When the prover is honest, i.e., the prover sends $|G\rangle^{\otimes Nk+m+1}$ to the verifier, the verifier accepts him with probability $1$, which is obvious from Eq.~(\ref{16}). This means that our verification protocol has the completeness. Hereafter, we consider the case where $\epsilon=1/(4Nk^{2/7})$, $m\ge2N^3k^{18/7}\log{2}$, and $k\ge (4N)^7$ are satisfied. In this case, the following theorem holds.

\begin{theorem}[Soundness]
\label{soundness_3}
If the verifier accepts the prover, the state $\rho_{\rm tgt}$ of the target register satisfies 
\begin{eqnarray*}
\langle G|\rho_{\rm tgt}|G\rangle\ge 1-k^{-1/7}
\end{eqnarray*}
with a probability larger than $1-k^{-1/7}$.
\end{theorem}
{\it Proof.} Let $\Pi^\perp$ be the $N$-qubit projector $I^{\otimes N}-|G\rangle\langle G|$, and $T$ be the POVM element corresponding to the event where the verifier accepts the prover. We can show that for any $N$-qubit state $\rho$,
\begin{eqnarray}
\label{bound3}
{\rm Tr}[(T\otimes\Pi^\perp)\rho^{\otimes Nk+1}]\le\cfrac{1}{2k^{2/7}}.
\end{eqnarray}
Its proof is given later. Because of the quantum de Finetti theorem (for the fully one-way LOCC norm)~\cite{[LS15]} and Eq.~(\ref{bound3}),
\begin{eqnarray*}
{\rm Tr}[(T\otimes\Pi^\perp)\rho'_B]&\le& {\rm Tr}\left[(T\otimes\Pi^\perp)\int d\mu\rho^{\otimes Nk+1}\right]\ \ \ \ \ \\
&&+\cfrac{1}{2}\sqrt{\cfrac{2N^3k^2\log{2}}{m}}\\
&\le&\cfrac{1}{2k^{2/7}}+\cfrac{1}{2k^{2/7}}=\cfrac{1}{k^{2/7}}.
\end{eqnarray*}
Here, $\mu$ is a probability measure on $\rho$. We have 
\begin{eqnarray*}
{\rm Tr}[(T\otimes\Pi^\perp)\rho'_B]={\rm Tr}[(T\otimes I)\rho'_B]{\rm Tr}[\Pi^\perp\rho_{\rm tgt}].
\end{eqnarray*}
Therefore, if 
\begin{eqnarray*}
{\rm Tr}[\Pi^\perp\rho_{\rm tgt}]>k^{-1/7},
\end{eqnarray*}
then 
\begin{eqnarray*}
{\rm Tr}[(T\otimes I)\rho'_B]<k^{-1/7}.
\end{eqnarray*}
This means that if the verifier accepts the prover, 
\begin{eqnarray*}
\langle G|\rho_{\rm tgt}|G\rangle\ge 1-k^{-1/7}
\end{eqnarray*}
with a probability larger than $1-k^{-1/7}$.

To complete the proof, we show Eq.~(\ref{bound3}). First, we consider the case where ${\rm Tr}[g_i\rho]\ge 1-4\epsilon$ for all $i$. Because of the union bound,
\begin{eqnarray*}
{\rm Tr}[\Pi^\perp\rho]&=&1-{\rm Tr}\left[\prod_{i=1}^N\cfrac{I^{\otimes N}+g_i}{2}\rho\right]\\
&\le&\sum_{i=1}^N\left(1-{\rm Tr}\left[\cfrac{I^{\otimes N}+g_i}{2}\rho\right]\right)\\
&\le&2N\epsilon\\
&=&\cfrac{1}{2k^{2/7}}.
\end{eqnarray*}
Therefore,
\begin{eqnarray}
\nonumber
{\rm Tr}[(T\otimes\Pi^\perp)\rho^{\otimes Nk+1}]&=&{\rm Tr}[T\rho^{\otimes Nk}]{\rm Tr}[\Pi^\perp\rho]\\
\label{poly}
&\le&\cfrac{1}{2k^{2/7}}.
\end{eqnarray}
Next, we consider the case where ${\rm Tr}[g_i\rho]< 1-4\epsilon$ is satisfied for at least one $i$. In this case, for the $i'$ that satisfies ${\rm Tr}[g_{i'}\rho]< 1-4\epsilon$, 
\begin{eqnarray*}
p_{{\rm pass}}(i')=\cfrac{1+{\rm Tr}[g_{i'}\rho]}{2}<1-2\epsilon.
\end{eqnarray*}
Therefore, because of the Hoeffding inequality,
\begin{eqnarray*}
{\rm Tr}[(T\otimes I)\rho^{\otimes Nk+1}]&\le&{\rm Pr}\left[\cfrac{K_{i'}}{k}\ge1-\epsilon\right]\\
&\le&{\rm Pr}\left[\cfrac{K_{i'}}{k}> p_{{\rm pass}}(i')+\epsilon\right]\\
&\le&e^{-2\epsilon^2k}\\
&=&e^{-k^{3/7}/(8N^2)}\\
&\le&e^{-2k^{1/7}}.
\end{eqnarray*}
Hence, 
\begin{eqnarray}
\nonumber
{\rm Tr}[(T\otimes\Pi^\perp)\rho^{\otimes Nk+1}]&=&{\rm Tr}[T\rho^{\otimes Nk}]{\rm Tr}[\Pi^\perp\rho]\\
\label{exp}
&\le&e^{-2k^{1/7}}.
\end{eqnarray}
From, Eqs.~(\ref{poly}) and (\ref{exp}),
\begin{eqnarray*}
{\rm Tr}[(T\otimes\Pi^\perp)\rho^{\otimes Nk+1}]&\le&{\rm max}\left(\cfrac{1}{2k^{2/7}},e^{-2k^{1/7}}\right)\\
&=&\cfrac{1}{2k^{2/7}}.
\end{eqnarray*}
\hspace{\fill}$\blacksquare$

\section{Applications}
In this section, we discuss applications of our
protocols to
the verification of quantum computational supremacy demonstrations with IQP circuits
and its variants, and verifiable blind quantum computing. 

First, we discuss the verification of quantum computational supremacy demonstrations with IQP circuits. An $N$-qubit IQP circuit is the following restricted quantum circuit:
\begin{itemize}
\item[(i)]
The initial state is $|0\rangle^{\otimes N}$.
\item[(ii)]
The $N$-qubit unitary $H^{\otimes N}DH^{\otimes N}$ is applied, 
where $H$ is the Hadamard gate, and $D$ is 
a quantum circuit consisting of a polynomial number
of $Z$-diagonal gates, such as $Z$, $CZ$, and $e^{i\theta Z}$.
\item[(iii)]
Finally, each qubit is measured in the computational basis.
\end{itemize}
The IQP model does not seem to be universal, but it is known that
the output probability distributions of
the IQP model cannot be classically efficiently sampled with a
constant multiplicative error unless
the polynomial-time hierarchy (PH) collapses to the third level~\cite{[BJS11]}
or the second level~\cite{[FKMNTT16]}.
Here, we say that a probability distribution $\{p_z\}_z$ is
sampled with a multiplicative error $\epsilon$ if
\begin{eqnarray*}
|p_z-q_z|\le \epsilon p_z
\end{eqnarray*}
for all $z$, where $q_z$ is the probability that the classical sampler
outputs $z$.

Recently, Bremner, Montanaro, and Shepherd~\cite{[BMS16]} have shown that,
assuming a certain unproven conjecture,
the no-go result can be generalized to the $l_1$-norm error sampling,
which is more realistic.
Here, we say that a probability distribution $\{p_z\}_z$ is
sampled with an $l_1$-norm error $\epsilon$ if
\begin{eqnarray*}
\sum_z|p_z-q_z|\le \epsilon,
\end{eqnarray*}
where $q_z$ is the probability that the classical sampler outputs $z$.
More precisely, they have shown
the following theorem.
\begin{theorem}[Ref.~\cite{[BMS16]}]
\label{Bremner}
Assume the below conjecture is true. If it is possible to classically sample from the output probability distribution of any IQP circuit in polynomial time, up to an error of $1/192$ in $l_1$ norm, then there is a ${BPP}^{NP}$ algorithm to solve any problem in $P^{\#P}$. Hence, the PH would collapse to its third level.
\end{theorem}
\begin{conjecture}[Ref.~\cite{[BMS16]}]
\label{conjecture}
Let $f:\{0, 1\}^N\rightarrow\{0, 1\}$ be a uniformly random degree-$3$ polynomial over $\mathbb{F}_2$. Then, it is $\#$P-hard to approximate $({\rm gap}(f)/2^N)^2$ up to a multiplicative error of $1/4+o(1)$ for a $1/24$ fraction of polynomials $f$. Here, ${\rm gap}(f)\equiv|\{x:f(x)=0\}|-|\{x:f(x)=1\}|$. 
\end{conjecture}
Here, complexity classes BPP, NP, P, and $\#$P are abbreviations of bounded-error probabilistic polynomial-time, nondeterministic polynomial-time, polynomial-time, and sharp-P, respectively.

Importantly,
the theorem holds for the IQP model that uses
only $Z$, $CZ$, and $CCZ$ gates, where $CCZ$ is the
controlled-controlled-$Z$ gate defined as
\begin{eqnarray*}
CCZ=I^{\otimes 3}-2|111\rangle\langle111|.
\end{eqnarray*}
The theorem therefore shows the hardness for the sampling
of the probability distribution
of the $X$-basis measurement outcomes on hypergraph states.
In other words, if the verifier generates a hypergraph state in his or her laboratory or
receives it from a remote server,
the verifier can demonstrate the quantum computational supremacy.
However, one problem is that what the verifier receives deviates from the
ideal hypergraph state because of the experimental
imperfections or the server's dishonesty.
The verifier therefore has to verify the correctness of the state,
where the verification task becomes important.

In Ref.~\cite{[BMS16]},
all gates, $Z$, $CZ$, and $CCZ$, are applied uniformly random.
The anticoncentration lemma,
which is essential for their proof,
is satisfied when $Z$ and $CZ$ gates are applied uniformly random,
but Conjecture~\ref{conjecture},
which is often called ``average case vs worst case hardness conjecture,"
seems to be more plausible when the application of
$CCZ$ gates is also uniformly random. 
In other words, the hypergraph states generated by the 
IQP circuits of Ref.~\cite{[BMS16]}
can have high connectivity.

Our third protocol can verify such hypergraph states with high connectivity.
From Theorem~\ref{soundness_3}, we can guarantee 
that 
\begin{eqnarray*}
\frac{1}{2}\Big\|\rho_{\rm tgt}-|G\rangle\langle G|\Big\|\le
\sqrt{1-\langle G|\rho_{\rm tgt}|G\rangle}\le\frac{1}{{\rm poly}(k)},
\end{eqnarray*}
which means
\begin{eqnarray*}
\frac{1}{2}\sum_x|{\rm Tr}[M_x\rho_{\rm tgt}]-
\langle G|M_x|G\rangle|
&\le&
\frac{1}{2}\Big\|\rho_{\rm tgt}-|G\rangle\langle G|\Big\|\\
&\le&
\frac{1}{{\rm poly}(k)}
\end{eqnarray*}
for any POVM $\{M_x\}_x$. Here, $\|\cdot\|$ is the trace norm. In particular, if we take the
POVM as the $X$-basis measurements,
\begin{eqnarray*}
\sum_z|p_z-p'_z|\le\frac{1}{{\rm poly}(k)},
\end{eqnarray*}
where $p_z$ is the probability of obtaining the outcome $z(\in\{0,1\}^N)$
when $|G\rangle$ is measured in the $X$ bases,
and
$p_z'$ is the probability of obtaining the outcome $z$
when $\rho_{\rm tgt}$ is measured in the $X$ bases:
\begin{eqnarray*}
p_z&=&|\langle z|H^{\otimes N}|G\rangle|^2,\\
p_z'&=&\langle z|H^{\otimes N}\rho_{\rm tgt}H^{\otimes N}|z\rangle.
\end{eqnarray*}
Assume that $\{p_z'\}_z$
is classically
efficiently sampled with the $l_1$-norm error $1/193$:
\begin{eqnarray*}
\sum_z|p'_z-q_z|\le\frac{1}{193},
\end{eqnarray*}
where $q_z$ is the probability that a classical sampler outputs $z$.
Then,
\begin{eqnarray*}
\sum_z|p_z-q_z|&\le&\sum_z|p_z-p'_z|+\sum_z|p'_z-q_z|\\
&\le&\frac{1}{
{\rm poly}(k)}+\frac{1}{193}\le\frac{1}{192},
\end{eqnarray*}
which causes the collapse of the PH
according to Theorem~\ref{Bremner}.
In conclusion, the probability distribution of the
$X$-basis measurement outcomes on the verified state
$\rho_{\rm tgt}$ through our third protocol
cannot be classically efficiently sampled
with the $l_1$-norm error. Similarly, our third protocol can also be used to verify variants of the 
IQP model such as those introduced in Refs.~\cite{[TT16],[GWD17],[VHSRE17],[MSM17],[BGK17]}.

Recently, several other verification protocols for IQP circuits have also been proposed. For example, Hangleiter {\it et al.} have proposed a polynomial-time verification protocol~\cite{[HKSE17]} using the Feynman-Kitaev history state~\cite{[KSV02],[F86]}
\begin{eqnarray*}
\cfrac{1}{\sqrt{L+1}}\sum_{t=0}^L\left(\prod_{i=0}^tU_i|\phi_0\rangle\right)\otimes|t\rangle
\end{eqnarray*}
corresponding to the quantum circuit $\prod_{i=1}^LU_i$ with an initial state $|\phi_0\rangle$, where $U_0=I$. In their protocol, the prover sends the Feynman-Kitaev history state to the verifier. Since the Feynman-Kitaev history state is, in general, more complicated than the mere output state $(\prod_{i=1}^LU_i)|\phi_0\rangle$, their protocol is more demanding for the prover than ours. Their protocol is also more demanding for the verifier because multiqubit measurements are necessary for the verifier. Moreover, their protocol assumes the i.i.d. property of samples unlike ours. Miller {\it et al.} have proposed another polynomial-time verification protocol for IQP circuits~\cite{[MSM17]}. Although the prover in their protocol only has to generate hypergraph states like our protocol, their protocol also assumes the i.i.d. property of samples. A verification protocol proposed in Ref.~\cite{[MTH17]} does not assume any i.i.d. property of samples, but the protocol cannot be used for hypergraph states with high connectivity because exponentially many quantum states are required to distinguish the probability of passing a test from 1/2,
which means that no polynomial-time verification is possible. Accordingly, this protocol cannot be used to verify the Bremner-Montanaro-Shepherd-type IQP circuits of Ref.~\cite{[BMS16]}.

As another application, our verification protocol for hypergraph states can also be used to construct a verifiable blind quantum computing protocol in a similar way to Ref.~\cite{[MTH17]}. Since the Union Jack state~\cite{[MM16]}, which is a hypergraph state, is a universal resource state for MBQC with only adaptive single-qubit measurements of Pauli operators, the client is required to perform only single-qubit Pauli measurements.

\section{Discussion}
We have seen that if the honest prover sends the correct state
to the verifier, the verifier accepts it with high probability.
However, in reality, it is not easy for the verifier
to receive the perfectly ideal state: Imperfections in the prover's machine 
and noises in the channel from the prover to the verifier change the state
even if the prover is honest.
In this section, we point out that even if the state is slightly
deviated from the ideal one, the verifier still accepts with high probability.
In other words, our protocols are robust to some extent.
We also discuss possibilities of using the quantum error correction.

To understand our argument, let us consider a simple example. 
Assume that the verifier receives the slightly deviated state
\begin{eqnarray}
\left[\left(1-\epsilon'\right)|G\rangle\langle G|+\epsilon' \eta\right]^{\otimes Nk+m+1} 
\label{deviate}
\end{eqnarray}
{\red instead} of $|G\rangle\langle G|^{\otimes Nk+m+1}$,
where $0<\epsilon'<1$, $|G\rangle$ is the ideal hypergraph state, and
$\eta$ is any state.
The trace distance between the deviated state and the ideal state is
\begin{eqnarray*}
\frac{1}{2}\Big\|
(1-\epsilon')|G\rangle\langle G|+\epsilon' \eta-
|G\rangle\langle G|\Big\| 
\le \sqrt{\epsilon'},
\end{eqnarray*}
and therefore, if $\epsilon'=O(1/{\rm poly})$, the deviated state
is still useful for the quantum computational supremacy demonstration.
This means that the deviated state should also be accepted by the verifier with
high probability.
In fact, our protocol accepts it with high probability. From Eq.(\ref{16}),
\begin{eqnarray*}
p_{\rm pass}(i)&=&\frac{1+{\rm Tr}[\rho g_i]}{2}\\
&=&1-\cfrac{\epsilon'}{2}\left(1-{\rm Tr}[\eta g_i]\right)
\end{eqnarray*}
for each $i=1,2,...,N$. Therefore,
the probability that the verifier accepts the deviated state is
\begin{eqnarray*}
{\rm Pr}[\mbox{verifier accepts}]
&=&{\rm Pr}\left[\bigwedge_{i=1}^N\left(\frac{K_i}{k}\ge1-\epsilon\right)\right]\\
&\ge&1-\sum_{i=1}^N {\rm Pr}\Big[\frac{K_i}{k}<1-\epsilon\Big]\\
&\ge&1-Ne^{-2(\epsilon'-\epsilon)^2k}.
\end{eqnarray*}
Since $k\ge (4N^7)$, if $\epsilon'-\epsilon=O(N^{-3})$, $1-Ne^{-2(\epsilon'-\epsilon)^2k}$ approaches $1$ asymptotically.

For simplicity, in the above example,
we have considered the tensor product of
the same states, Eq.~(\ref{deviate}), but 
it is easy to confirm that
a similar argument holds even if the tensor product state is replaced
with a slightly entangled state.

In this way, we have seen that our protocols are robust to some extent.
However, we have to mention that our protocols are not perfectly fault tolerant.
For example, let us consider the state 
\begin{eqnarray*}
(Z_1\otimes I^{\otimes N-1})|G\rangle,
\end{eqnarray*}
where only the first qubit of the ideal hypergraph state is phase flipped.
Such a state should also be accepted with high probability
because such a tiny error can be easily corrected with 
a quantum error correction; thus, the corrected state
is a useful resource state for the verifier. 
However, it is also easy to check that our protocols cannot
accept such a state with high probability because such a state is stabilized by $-g_1$, where $g_1$ is the first stabilizer of the ideal state $|G\rangle$.

A solution to the problem is to ask the prover to
send the encoded version of $|G\rangle$ with the 
Calderbank-Shor-Steane (CSS) code~\cite{CS96,S96}. 
(This means that the prover encodes each qubit of $|G\rangle$
into a logical qubit with the CSS code.)
A great advantage of our protocols is that only Pauli measurements
are required for the verifier. Since in the CSS code logical Pauli
measurements can be done with the transversal physical Pauli measurements,
the verifier can do the verification and the syndrome measurements
with only physical single-qubit 
Pauli measurements; i.e., no entangling gate is required for the verifier.

In Refs.~\cite{FH17,[MFN16]}, 
more elaborated methods have been proposed.
{\red Instead} of physically encoding states, the prover sends special states,
such as the Raussendorf-Harrington-Goyal (RHG) topological graph 
state~\cite{RHG07},
so that the verifier can do the topological quantum error correction
with only physical single-qubit Pauli measurements.
Unfortunately, such a scheme is known only for graph states,
and at this moment, we do not know how to generalize it
to hypergraph states.
If a similar scheme is found for hypergraph states, we can apply it to our verification protocols so that the
verifier can accept a broad class of deviated but topologically
correctable states with high probability.

{\red With respect to other verification protocols for ground states of Hamiltonians and output states of quantum circuits, a similar argument holds from Eqs.~(\ref{pass_ground}) and (\ref{pass_circuit}).}

\section{Conclusion}
In this paper, we have proposed verification protocols for 
ground states of Hamiltonians, quantum states generated by a 
certain class of quantum circuits, and all polynomial-time-generated 
hypergraph states. As applications of our verification protocols, 
we have considered the verification of IQP circuits and its variants, 
and verifiable blind quantum computing. 

As an outlook, let us finally provide several open problems.
First, our verification protocol for ground states of Hamiltonians
requires knowledge of, for example, the ground energy and energy gap.
It is, in general, QMA-hard to know these quantities, and therefore,
it is an important open problem {\red whether or not} a protocol
that does not use this knowledge {\red exists}. {\red If it exists, it is desirable to invent such a protocol.} More precisely, it is unknown whether or not conditions (i)--(iii) for the Hamiltonian in Sec. II can be relaxed. Related to this open problem, it is interesting to consider the physical relevance of the conditions. This is also the case for our second verification protocol, namely, the verification protocol for quantum circuits. It is an important open problem to find physical meaning of conditions (i)--(iii) for the circuit and to relax these conditions.

Second, it would be useful to consider verification protocols for 
other quantum states such as weighted graph states,
\begin{eqnarray*}
\left(\prod_{(i,j)\in E}e^{i\theta_{ij}Z_iZ_j}\right)|+\rangle^{\otimes N},
\end{eqnarray*}
and 
higher-dimensional quantum states including the
continuous-variable ones.
Here, $E$ is a set of edges, and $\theta_{ij}\in\mathbb{R}$. 
Those
states are important resources in 
quantum information and condensed matter physics~\cite{[GE07],[HCDB07]}. 

Finally, with respect to the verification of 
quantum computational supremacy demonstrations, 
it would be interesting to explore good verification protocols for subuniversal circuits other than the IQP, such as
the DQC1 model~\cite{[KL98],[MFF14],[FKMNTT16],[M17]}, 
the boson sampling model~\cite{[AA13]}, 
the depth-four model~\cite{[TD04]}, 
the Fourier sampling model~\cite{[FU15]}, 
and the conjugated Clifford model~\cite{[BFK17]}.

\section*{ACKNOWLEDGEMENTS}

We thank anonymous referees for valuable comments. Y.T. is supported by the Program for Leading Graduate Schools: Interactive Materials Science Cadet Program and JSPS Grant-in-Aid for JSPS Research Fellow No. JP17J03503. T.M. is supported by JST ACT-I No. JPMJPR16UP, the JSPS Grant-in-Aid for Young Scientists (B) No. 17K12637, and JST, PRESTO, No. JPMJPR176A.


\begin{thebibliography}{99}
\bibitem{FN1}When resource states are used to solve a decision problem in nondeterministic polynomial-time (NP), we do not have to verify the correctness of the resource states because the correctness of outputs can be classically verified. However, since the relationship between bounded-error quantum polynomial-time (BQP) and NP is not well known, such a classical verification method cannot always be used for decision problems in BQP.
\bibitem{[FN3]} From the model theoretical perspective, it is preferable to call these procedures ``certification." However, in this paper, we use ``verification" instead because we are also interested in the interactive proof setup such as verifiable blind quantum computing.
\bibitem{FN2}Not all CPTP maps can be implemented in quantum polynomial time. However, in cloud quantum computing, i.e., verifiable blind quantum computing~\cite{[MF13],[HM15]}, the security is guaranteed for any CPTP map performed by the malicious server. Therefore, we assume that $\mathcal{E}$ is any CPTP map. This situation is similar to quantum key distribution, where an eavesdropper is assumed to be able to perform any CPTP map as an attack.
\bibitem{mixed}
Here, we assume that the correct state to be verified is pure.
If it is not pure, the SWAP test
is not enough. Actually, distinguishing mixed states
is quantum statistical zero-knowledge (QSZK)-complete.
\end{thebibliography}
\end{document}